\documentclass[aps,prl,superscriptaddress, nofootinbib,reprint,nolongbibliography]{revtex4-2}

\usepackage{graphicx}
\usepackage{makecell}
\usepackage{lineno}
\usepackage{amsmath}
\usepackage[table]{xcolor}
\usepackage{ulem}
\usepackage{physics}

\begin{document}

\title{Sympathetic cooling and slowing of molecules with Rydberg atoms}

\author{Chi Zhang}
\email[]{chizhang@caltech.edu}
\affiliation{Division of Physics, Mathematics, and Astronomy, California Institute of Technology, Pasadena, CA 91125, USA}

\author{Seth T. Rittenhouse}
\affiliation{Department of Physics, the United States Naval Academy, Annapolis, Maryland 21402, USA}

\author{Timur V. Tscherbul}
\affiliation{Department of Physics, University of Nevada, Reno, Nevada 89557, USA}

\author{H. R. Sadeghpour}
\affiliation{ITAMP, Center for Astrophysics $|$ Harvard $\&$ Smithsonian Cambridge, Massachusetts 02138, USA}

\author{Nicholas R. Hutzler}
\affiliation{Division of Physics, Mathematics, and Astronomy, California Institute of Technology, Pasadena, CA 91125, USA}


\begin{abstract}

We propose to sympathetically slow and cool polar molecules in a cold, low-density beam using laser-cooled Rydberg atoms. The elastic collision cross sections between molecules and Rydberg atoms are large enough to efficiently thermalize the molecules even in a low density environment. Molecules traveling at 100 m/s can be stopped in under 30 collisions with little inelastic loss. Our method does not require photon scattering from the molecules and can be generically applied to complex species for applications in precision measurement, quantum information science, and controlled chemistry.

\end{abstract}

\maketitle
Cold and trapped molecules are unique quantum systems for a multitude of applications ranging from table-top search of new physics beyond the Standard Model \cite{Safronova2018,Hudson2011,Baron2014,Andreev2018,Roussy2023,Truppe2013,Leung2023} to quantum information processing \cite{Holland2022,Bao2022,Lin2020,Park2017,Gregory2021,Burchesky2021}, which benefit from rich internal molecular structures. Recent measurements of the electron electric dipole moment (EDM) \cite{Roussy2023,Andreev2018}, which rely on the strong internal electric field and the high-level control of molecule orientation, have constrained charge-parity violating new physics to $\gtrsim 50~\mathrm{TeV}$ energy scales. In addition, the molecular rotational structure provides tunable long-range dipole-dipole interaction in ground electronic states \cite{Blackmore2018}, and may accommodate one error-corrected qubit in each molecule \cite{Albert2020}, thereby significantly reducing the number of physical qubits in a quantum information processor.

Laser cooling of molecules \cite{Barry2014,Truppe2017b,Cheuk2018,Caldwell2019,Ding2020,Langin2021} and assembling ultracold atoms \cite{Ni2008,DeMarco2019} are two of the main pathways to trapping molecules in the quantum regime, key to the next generations of new physics searches \cite{Alauze2021,Fitch2020b,Augenbraun2020,Kozyryev2017b,Isaev2010}, long-lived qubits and high-fidelity quantum gates. However, both methods require specific molecular structures, which limit the choice of molecules.  Furthermore, laser cooling needs $10^4 - 10^5$ photon scattering events and thus demands high precision spectroscopy, which is challenging and time-consuming for many heavy-atom containing \cite{Zhang2022,Persinger2022} or large polyatomic molecules \cite{Augenbrau2023,Mitra2022}.

Molecules can also be cooled by collisions or interactions with another species. For example, in a cryogenic buffer gas beam (CBGB) \cite{Hutzler2012}, cold helium gas thermalizes any molecular species to a few Kelvin temperature.  Once in a trap, the molecules can be further cooled by opto-electrical Sisyphus cooling~\cite{Prehn2016} or via sympathetic cooling with laser-cooled atoms~ \cite{Lim2015,Jurgilas2021,Rugango2015,Tscherbul2011,Morita2017,Son2020}. It would be advantageous to use laser-cooled species to sympathetically cool species in lower-density environments, such as in beams, since loading complex, reactive species into traps with sufficient atom and molecular density for efficient sympathetic cooling is often a challenge. However, due to the substantially lower density, collisions between molecules and ground state atoms are too rare for efficient cooling in a beam. 

Here, we propose a method to slow and cool polar molecules from a CBGB, load them into a trap, and cool to ultracold temperatures, without photon scattering from the molecule. The method relies on enhancing elastic collision cross sections between molecules and atoms via atomic Rydberg excitation, and thermalizing the molecules to laser cooled atoms.
The sequence begins with a CBGB of both the molecules and atoms. Next, laser cool and slow the atoms while exciting a fraction of them to Rydberg states. The Rydberg atoms, because of their large size, collide with the molecules and bring them into approximate equilibrium with the atom velocity distribution. Once the molecules and atoms are both stopped, a position-dependent Rydberg excitation in an applied external field gradient provides collisions and compresses the molecule density for loading into a trap.

Highly excited Rydberg atoms are orders of magnitude larger than ground state atoms in size, and this can compensate for the lower density and increase the collision rate. Early studies of molecule-Rydberg atom collisions at high collision energies (compared to the molecule-Rydberg atom interaction energies) measured vanishing cross sections at room temperature \cite{Stebbings1983,Gallagher1994}. 
At cold temperatures, however, the collision energy becomes smaller or comparable to the molecule-Rydberg atom interaction energy. 
This leads to cross sections as large as the size of the Rydberg wavefunction, which is around six orders of magnitude larger than the ground states for a typical Rydberg state, and thus will enhance the collision rate for efficient sympathetic cooling.

\begin{figure*}
	\includegraphics[width=0.8\textwidth]{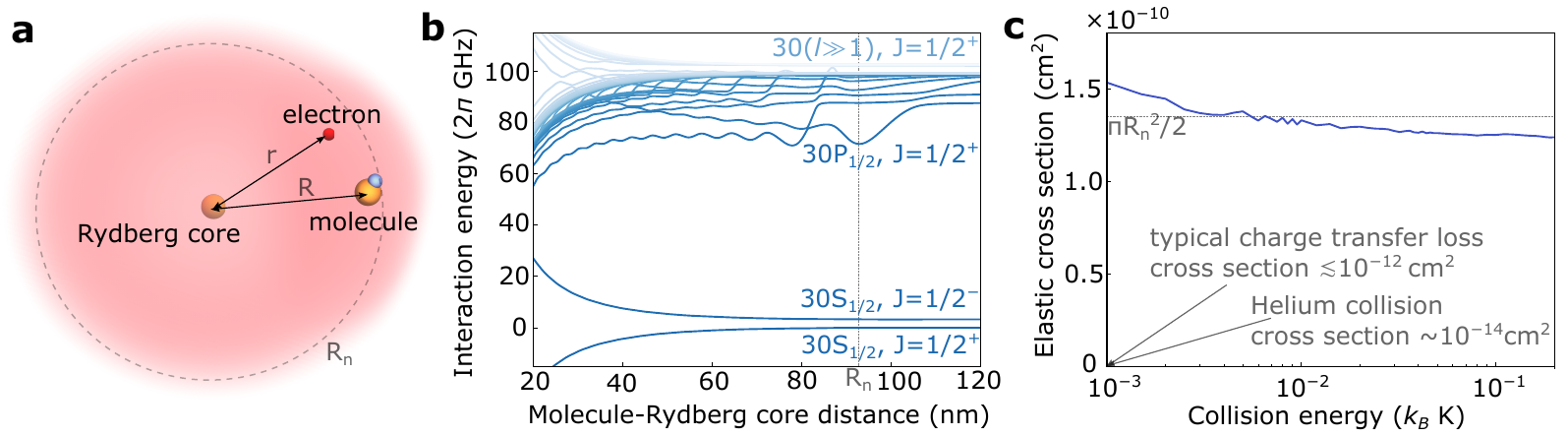}
	\caption{Molecule-Rydberg atom interaction and scattering cross section. (a) A molecule inside a Rydberg atom. $R$ is the distance between the molecule and the core of the atom, $r$ is the position of the electron relative to the core, and $R_n$ is the classical radius of the Rydberg state. The red background indicates the Rydberg wave function, which is slightly dressed by the molecule. (b) Interaction potential between the CH molecule and the Li Rydberg atom. The state labels indicate the asymptotic states when the molecule is far away from the Rydberg atom. When they are close, all atom-molecule pair states are hybridized. The dashed vertical line marks the classic radius of the Li($30S_{1/2}$) state. When the pair state density is high near the degenerate manifold of high angular momentum Rydberg states, the interaction potential depth is larger than $\sim 10~\mathrm{GHz}$. (c) The scattering cross section for the potential of the $30S_{1/2},J=1/2^-$ pair state, compared with the half of the cross section of a hard sphere with radius $R_n$ ($\pi R_n^2$).}
	\label{Fig1}
\end{figure*}

Our method has two requirements, which we will show to be generic for polar molecules. First, we require a mean free path which is short compared to the size of the atomic cloud, so that the molecules are thermalized before diffusing out of the cloud. This requires a deep atom-molecule interaction potential that is comparable to or larger than the collision energy in a CBGB ($\sim 10~\mathrm{m/s}$ in the moving frame, which corresponds to $\sim h \times 10~\mathrm{GHz}$ for heavy molecules of $\sim 200$ atomic mass units, where $h$ is Planck's constant). Second, we require that the loss probability during collisions is not too high. We will show that fewer than $\sim 30$ collisions are needed to bring a molecule from a few Kelvin and $\sim 100~\mathrm{m/s}$ forward velocity to ultracold temperature in a trap. This requirement is much less stringent than those imposed by the previous proposals of sympathetic cooling \cite{Lim2015,Jurgilas2021,Rugango2015,Tscherbul2011,Morita2017}, and we shall show that the loss probability per collision is sufficiently low in the general case.

When a molecule is outside of a Rydberg atom, the dominant interaction is the dipole-dipole interaction \cite{Urban2009}. This interaction has been used to detect and manipulate the state of the molecule \cite{Patsch2022,Zhelyazkova2017,Jarisch2018}, and has been proposed for cooling trapped molecules in $\sim \mathrm{mK}$ temperature regime \cite{Huber2012,Zhao2012} and for entangling molecules in optical tweezers \cite{Zhang2022b,Wang2022,Kuznetsova2011}. However, the dipole-dipole interaction is $\ll h\times 100~\mathrm{MHz}\approx k_B \times 5~\mathrm{mK}$, where $k_B$ is Boltzmann's constant. This is negligible compared to the collision energy in the $\sim \mathrm{K}$ temperature regime, which is of interest in this work.

The picture changes drastically when the molecule is near the edge or inside the Rydberg wavefunction. As illustrated in Fig.~\ref{Fig1}(a), the separation between the molecule and the Rydberg core $R$ is comparable to or smaller than the electron-core distance $r$, and the electron can be very close to the molecule. As a result, the dominant perturbation of the molecule on the Rydberg atom comes from the charge-dipole interaction with the Rydberg electron, $H_I = ed/|R-r|^2$ with $d$ the molecule frame dipole moment. This interaction has been intensively studied in systems of ultralong range Rydberg molecules \cite{Guttridge2023,Rittenhouse2010,Rittenhouse2011, Shaffer2018,GonzalezFerez2015,AguileraFernandez2017,GonzalezFerez2021}. It couples and hence shifts the atom-molecule pair states strongly. The coupling strength is proportional to $d$ and the electron's probability density which scales as $(n^*)^{-3}$, with $n^*$ the effective principal quantum number.

In Fig.~\ref{Fig1}(b), we show the interaction potentials between the CH molecule ($d\approx 1.46~\mathrm{D}$) and the Li Rydberg atom as an example.  We choose this atom-molecule pair since the small dipole moment of the molecule enables accurate computations; however, this method generalizes to more complex species for which exact calculations might be impractical. For the near-degenerate manifold of high angular momentum states, the pair states are well hybridized and the potentials are $\sim 10~\mathrm{GHz}$ deep. The potentials for the S Rydberg states are weaker but still exceed GHz inside the Rydberg radius. For molecules with larger dipole moment, especially for $d\gtrsim 1.6~\mathrm{D}$ (Fermi-Teller critical dipole \cite{Fermi1947}), the non-degenerate pair states with very large energy differences are also well hybridized and the Rydberg electron becomes highly localized around the molecule. The potential energy curves become complex, however; on average, interaction potentials as deep as the quantum defects ($\gtrsim 20~\mathrm{GHz}$) can be expected. More details about the interaction between large-dipole molecules and Rydberg atoms
can be found in the Supplemental Material.

The range of the $\gtrsim \mathrm{GHz}$ potential is similar to the radius of the Rydberg wave function $R_n \approx 2(n^*)^2 a_0$, where $a_0$ is the Bohr radius. The relative collision velocity (in the center of mass frame) in a CBGB is $\lesssim 10~\mathrm{m/s}$, which roughly corresponds to $\lesssim~h\times0.5~\mathrm{GHz}$ collision energy for the CH molecule and the Li atom. This is much less than the potential depth, therefore the cross section is expected to be nearly the size of the potential, $\sim\pi R_n^2$. Even for heavier molecules with up to $\sim 200$ atomic mass units, the $\sim~h\times10~\mathrm{GHz}$ deep potentials are also sufficient for large cross sections. As an example, the scattering cross section of CH and Li is calculated and plotted in Fig.~\ref{Fig1}(c). For $n^* \approx 30$, the cross section is around four orders of magnitude larger than that of the helium atom, the most common buffer gas. This large cross section enables a short mean free path even in the dilute beam outside of the buffer gas cell. As we will show later, molecules at $\sim 100~\mathrm{m/s}$ initial velocity can be stopped and trapped by fewer than 30 collisions with decelerating Rydberg atoms of the same initial velocity.

\begin{figure}
	\includegraphics[width=0.5\textwidth]{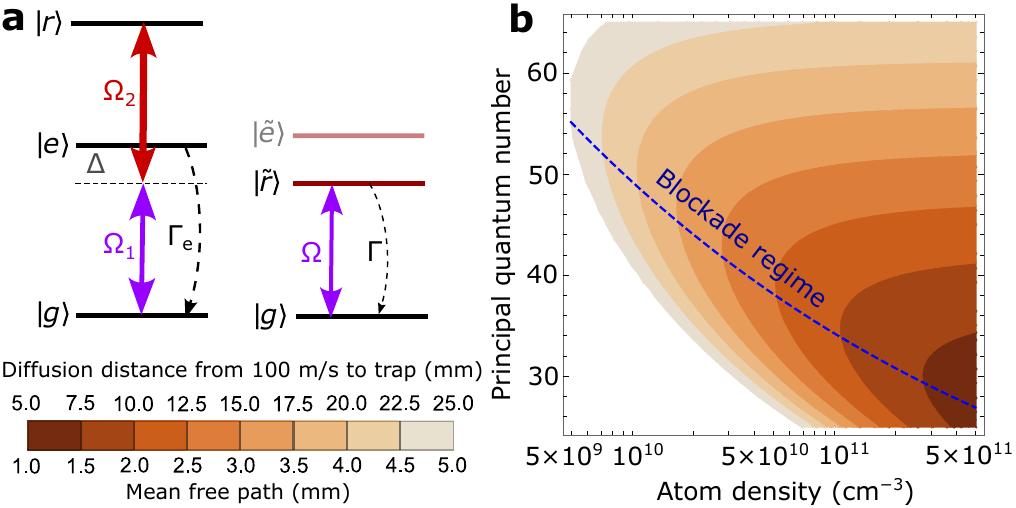}
	\caption{(a) Laser coupling scheme for Rydberg excitation and cooling. $\Omega_2$ (red arrow) couples $\ket{e} \leftrightarrow \ket{r}$ off-resonantly and $\Omega_1$ (purple arrow) couples $\ket{g}$ to the dressed Rydberg state $\ket{\tilde{r}}$ with an effective coupling strength $\Omega \approx 2\pi \times 1~\mathrm{MHz}$ (purple arrow on the right). (b) Mean free path of the collision, as well as the diffusion distance (in the atom cloud) to stop a molecule initially at 100~m/s, as a function of the total atom density (Rydberg and ground states) and principal quantum number, when the system is in a steady state with the laser couplings in (a) We use 25 as the number of collisions needed to stop the molecules, as described in the main text. The blue dashed line marks the parameters above which the blockade effect starts to limit the Rydberg population density.}
	\label{Fig2}
\end{figure}

\begin{figure}
    \includegraphics[width=0.5\textwidth]{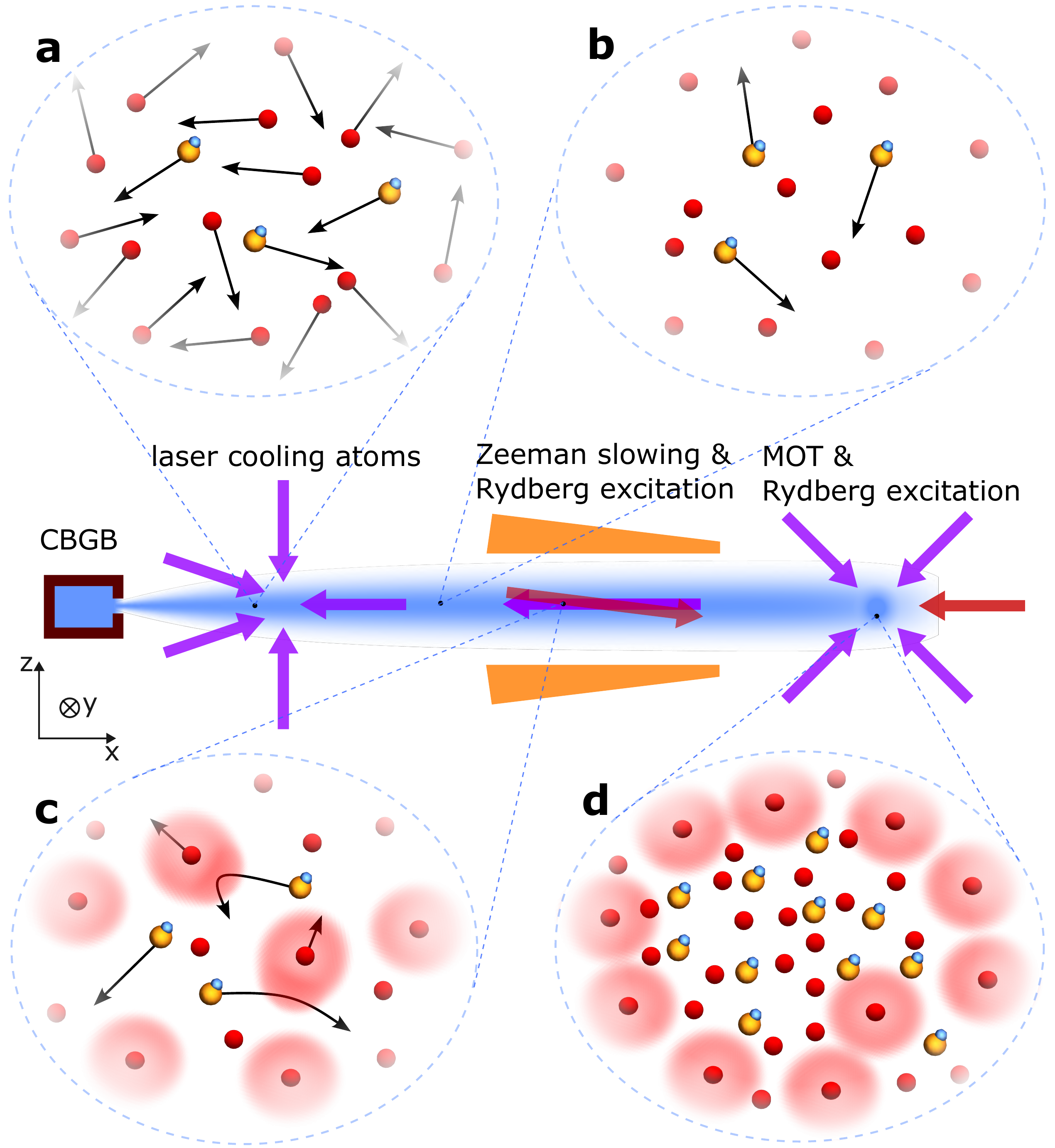}
	\caption{Proposed experimental setup and sequence. A beam (blue) of atoms and molecules comes out from a buffer gas cell (brown box). In (a), atoms and molecules are both relatively hot. Polarization gradient cooling is applied to the atoms in the moving frame. The frequencies of the beams along $x$ direction are shifted for cooling in the moving frame. Two beams in the $y$ direction are not shown in the figure. After cooling, in (b), atoms are cold but molecules are still hot. In the Zeeman slower (orange), atoms are excited to the dressed Rydberg state $\ket{\tilde{r}}$ by two counter-propagating beams. In (c), molecules collide and thermalize with the Rydberg atoms. After slowing, magneto-optical force and position dependent Rydberg excitation are applied to compress the density of the molecules and load them into a trap. (d) shows the magneto-optical trap where the molecules are pushed by the Rydberg atoms towards the center of the trap.}
	\label{Fig3}
\end{figure}

This enhancement of collision cross-section is general, and can be used for a variety of experimental goals. We now discuss a specific approach to implementing this method to slow, stop, trap, and cool molecules from a CBGB without laser cooling them. In a cloud of atoms, we consider the laser coupling scheme shown in Fig.~\ref{Fig2}(a) for Rydberg excitation and for Zeeman slowing of the atom. The excited state $\ket{e}$ and Rydberg state $\ket{r}$ are coupled by a blue detuned Rydberg laser with coupling strength $\Omega_2$ and detuning $\Delta$, the dressed eigenstate with dominant Rydberg character is $\ket{\tilde{r}}=\cos{\theta} \ket{r} - \sin{\theta} \ket{e}$, with the mixing angle $\theta$ given by $\tan{2 \theta} = -\Omega_2 /\Delta$. We consider $\Delta \not\gg \Omega_2 > \Gamma_e$, as the Rydberg state linewidths ($\lesssim 100~\mathrm{kHz}$) are much less than the low-lying excited state linewidth, therefore the linewidth of the dressed Rydberg state $\ket{\tilde{r}}$ is dominated by the contribution from $\ket{e}$, and it primarily decays to the ground state $\ket{g}$. For most alkali ($\Gamma_e \approx 2\pi \times 5~\mathrm{MHz}$) and alkaline earth ($\Gamma_e \approx 2\pi \times 30~\mathrm{MHz}$) atoms, the $\ket{\tilde{r}}$ state linewidth $\Gamma$ can be tuned to around $2\pi \times 1~\mathrm{MHz}$ by mixing a few percent or less $\ket{e}$ population. The effective coupling $\ket{g} \leftrightarrow \ket{\tilde{r}}$ can be tuned to around $\Omega \approx \Omega_1 \Omega_2 / 2\Delta = 2\pi \times 1~\mathrm{MHz}$, yielding a value for the saturation parameter $s =\frac{\Omega^2/2}{\Delta^2 + \Gamma^2/4} \approx 1$.

These parameters are suitable for most alkali and alkaline earth atoms. The time scale of an oscillation between $\ket{g} \leftrightarrow \ket{\tilde{r}}$ is much longer than the collision time ($\ll 0.1~\mathrm{\mu s}$,
the time during which the molecule is inside the Rydberg wavefunction). In the Zeeman slower, the two lasers $\Omega_1$ and $\Omega_2$ are counter-propagating and their Doppler shifts on $\ket{g} \leftrightarrow \ket{e}$ and $\ket{e} \leftrightarrow \ket{r}$ transitions are thus opposite. A position-dependent magnetic field shifts $\ket{e}$ (typically a $P$ state) differently from $\ket{g}$ and $\ket{r}$ ($S$ states) and thus can compensate the opposite Doppler shifts in $\ket{g} \leftrightarrow \ket{e}$ and $\ket{e} \leftrightarrow \ket{r}$ simultaneously. As a result, the Rydberg excitation scheme is not affected by the Zeeman slower for the target velocity class, and we can slow the atomic beam while maintaining a fixed Rydberg excitation fraction.

In steady state with these couplings, the mean free path, $l\approx(\sigma \rho)^{-1}$ where $\sigma$ is the elastic collision cross section and $\rho$ is the Rydberg atom density, is calculated and plotted in Fig.~\ref{Fig2}(b) as a function of total atom density (in both ground and Rydberg states) and principal quantum number. For low Rydberg states the dressed Rydberg population can reach 25\% (for $s=1$), while for high Rydberg states it is limited by the Rydberg blockade effect \cite{Urban2009,Gaetan2009}. The blue dashed line marks the parameters where Rydberg blockade starts to limit the density of atoms in Rydberg states. The collision cross section scales geometrically as $(n^*)^4$, until the interaction potential is too weak, but the probability of Rydberg excitation can be suppressed by Rydberg blockade, which scales as $(n^*)^7$. As a result, the optimal density and principal quantum number are near the blockade regime. 

The density of atoms made inside the buffer gas cell is typically 
$10^{13}$ or $10^{14}~\mathrm{cm}^{-3}$
\cite{Hutzler2012}. After exiting the cell, the density decreases rapidly because of expansion and collision with the buffer gas. We propose to first laser cool the atoms in all three dimensions to stop the expansion and keep the density high. The cooling needs to be in the moving frame to maintain the overlap between cooled atoms and uncooled molecules before subsequent Rydberg excitation (for collisions). At around $100~\mathrm{mm}$ outside the cell downstream the atom density is $10^{10}$ or $10^{11}~\mathrm{cm}^{-3}$ and the buffer gas collision rate is low enough for laser cooling. Normal Doppler laser cooling would not work since the atomic cloud is optically thick; instead, we consider polarization gradient cooling \cite{Devlin2016}. Cooling in the moving frame can be achieved by detuning the longitudinal cooling laser frequencies. Another option is to use the magneto-optical force to cool and compress the density of atoms in the two transverse directions and apply polarization gradient cooling (in the moving frame) in the forward direction.

Typical polarization gradient cooling can cool the atoms 
to $T_A < 100~\mathrm{\mu K}$, after about $1~\mathrm{cm}$ \cite{Devlin2016}, and then the atom density will stay at $10^{10} - 10^{11}~\mathrm{cm}^{-3}$. Subsequently, the cloud of ultracold atoms and uncooled molecules enters the Zeeman slower, where the atoms are excited to $\ket{\tilde{r}}$ for both slowing and collisions. The mean acceleration on the atoms is $a\approx \frac{s/2}{1+s} \frac{h \Gamma}{m_A \lambda}$, where $m_A$ is the mass of the atom and $\lambda$ is the wavelength of the transition. During the slowing process, molecules are also slowed by collisions with Rydberg atoms. In the moving frame of the ultracold atoms, the molecules have initial kinetic energies of $\sim 1~\mathrm{K}$ and are constantly accelerated by $a$, relative to the atoms until the cloud is stopped. The optimal mean free path $l$ can reach 2 to $3~\mathrm{mm}$ (see Fig.~\ref{Fig2}[b]). For atoms and molecules with similar masses, the expected molecule kinetic energy after one collision is $T_{M,f} \approx (T_{M,i} + T_A)/2$, where $T_{M,i}$ is the kinetic energy before the collision. We assume the atoms are not heated up by the collisions since there are typically orders of magnitude more atoms than molecules~\cite{Hutzler2012}. In steady state, the kinetic energy loss of the molecules in the collision $T_{M,i}-T_{M,f} \approx T_{M,f}$ is similar to the kinetic energy gained between two successive collisions by acceleration $m_M a l$, with $m_M$ the molecule mass (and $m_M \approx m_A$). Therefore, the collision energy $T_{M,i} \approx 2 m_M a l \approx \frac{s}{1+s} \frac{l}{\lambda} h \Gamma$. The collision energy is independent of the mass, and for typical parameters we choose ($s\approx 1$, $\Gamma \approx 2\pi \times 1~\mathrm{MHz}$) it is around $h\times 2~\mathrm{GHz}$, which is low enough for the interaction potential between most polar molecules and Rydberg states, though it can be further reduced by using smaller $s$ or $\Gamma$. The laser slowing distance of the entire cloud is $\frac{v^2}{2a}$, which is proportional to the atom mass, is $0.5~\mathrm{m}$ for heavy atoms with $\sim 200$ atomic mass unit and an initial velocity of $100~\mathrm{m/s}$. During the slowing process, on average a molecule collides $N \approx 25$ times with Rydberg atoms. This leads to a diffusion distance of $l \sqrt{N} \approx 1~\mathrm{cm}$ (shown in Fig.~\ref{Fig2}[b]), which is less than the typical size of the ultracold atomic cloud ($\sim 3~\mathrm{cm}$). On average, half of the molecules in the $\approx 1~\mathrm{cm}$ outer shell of the cloud may diffuse out. This results in a $\lesssim 30\%$ loss assuming uniform initial distribution. Although we assume the initial forward velocity is $\sim 100~\mathrm{m/s}$, we note that slow beams with $<50~\mathrm{m/s}$ from CBGB have been experimentally demonstrated \cite{Augenbrau2023,Augenbraun2021,Sawaoka2023}, this allows for a shorter Zeeman slower and thus less diffusion loss.

After Zeeman slowing, the molecules can be loaded into a magnetic trap \cite{McCarron2018,Jurgilas2021}, or any other trap which is sufficiently deep. If further cooling is needed, the molecules could be directly laser-cooled if they can scatter a sufficient number of photons, though now with less stringent requirements on vibrational closure due to the fact that they are now already stopped. 

Alternatively we propose a method to capture and cool both the atoms and molecules with magneto-optical trap (MOT) but without scattering photons from the molecules. Atoms are excited to $\ket{\tilde{r}}$ (see Fig.~\ref{Fig2}) on the edge of the MOT, which provides a position-dependent force to confine the atoms. Molecules can collide with the shell of Rydberg atoms ($\ket{\tilde{r}}$) so that they are confined and further thermalized. In the meantime, as we slowly vary the frequencies of the lasers and shrink the Rydberg shell, the molecular density is compressed.

After $N$ more collisions, the molecules have temperature $T_M(N) \approx T_A + T_M(0) e^{-N/2}$, where $T_M(0)$ is the molecule temperature after Zeeman slowing. Within seven collisions, the molecules are thermalized to $\lesssim 1~\mathrm{mK}$. The shell thickness is determined by the linewidth of $\ket{\tilde{r}}$ and the magnetic field gradient. We need a large MOT \cite{Camara2014} with a thick shell of Rydberg atoms, for instance $d \approx 1~\mathrm{cm}$. The probability of molecules inside the shell diffusing out during the compression is $1-e^{-\sqrt{N} l/d}/2 \approx 20\%$. After the compression, the molecules can be loaded into an optical dipole, magnetic, or other trap.

By leveraging the large elastic collision rate between molecules and Rydberg atoms, we have shown that we can slow, trap, and cool molecules to ultracold temperatures without scattering photons from the molecule.  However, there can be unwanted
side-effects of \textit{inelastic} collisions, including rotational
state changes and charge transfer.  We argue that neither of these are likely to be a major limitation in the general case. 

First, we consider rotational state changes.  Like elastic collisions with helium \cite{Hutzler2012}, we expect rotational states to thermalize upon collisions with non-negligible probability, resulting in a thermal rotational distribution, similar to the initial, thermalized, buffer gas cell rotational distribution. The collisions are unlikely to excite high rotational states $(N\gtrsim 5)$ because of their high energies, and the interaction potential and the cross section are highly independent of rotational state for low $N$, therefore all rotational states should be cooled. However, the rotational population may have an equivalent temperature of a few K, though that can be compressed by ``rotational cooling'' optical pumping methods \cite{Ho2020} with a few photon scatters and negligible heating, or simply used as-is.

Second, there could be charge transfer collisions \cite{Markson2016}, whereby the Rydberg electron migrates from the atom to the molecule. These collisions have been studied both theoretically and experimentally \cite{Desfranois1994,Matsuzawa1975,Hamamda2015,Yu2012,Lebedev2013,Qian2019}, and have been observed only for molecules with $d >2.5~\mathrm{D}$ dipole moments. Furthermore, the cross section depends on the principal quantum number and normally has a peak. At several principal quantum numbers away from the peak, the charge transfer loss cross section is typically $\ll 10^{-12} \mathrm{cm}^2$, more than two orders of magnitude smaller than the elastic cross section. As a result, the inelastic collisions should have negligible loss effects, and more than half of the molecules from the $100~\mathrm{m/s}$ distribution should survive in an ultracold trap, making Rydberg atom sympathetic cooling more efficient than laser cooling with only a $10^{-5}$ photon-scattering leakage.

In summary, we have shown that polar molecules can be slowed and cooled by the collisions with laser cooled Rydberg atoms. Our method does not require scattering photons from the molecule, the interaction with Rydberg atom arises from the molecular dipole moment, and as a result, the method works generically for polar molecules. Our method provides a pathway into traps at ultracold temperatures for many molecules that do not have laser cycling transitions or that are hard to cool. This will significantly boost a host of applications using ultracold molecules, such as quantum information and table-top searches for new physics.

\begin{acknowledgements}
We thank Yi Zeng, Ashay Patel, Weibin Li, Lan Cheng, Tim Steimle, Zack Lasner, Mike Tarbutt, Rosario Gonzalez-Ferez and Gerard Higgins for helpful discussions. 
The work at the California Institute of Technology has been supported by Gordon and Betty Moore Foundation Award GBMF7947 and Alfred P. Sloan Foundation Award G-2019-12502. C.Z. acknowledges support from the David and Ellen Lee Postdoctoral Fellowship at Caltech. T.V.T. gratefully acknowledges support from the NSF CAREER award No. PHY-2045681. H.R.S. acknowledges support from the NSF through a grant for ITAMP at Harvard University.
\end{acknowledgements}

Correspondence and requests for materials should be addressed to C.Z.~(email: chizhang@caltech.edu).

\clearpage

\widetext
\begin{center}
\textbf{\large Supplemental Materials: Sympathetic cooling and slowing of molecules with Rydberg atoms}
\end{center}

\setcounter{equation}{0}
\setcounter{figure}{0}
\setcounter{table}{0}
\setcounter{page}{1}
\makeatletter
\renewcommand{\theequation}{S\arabic{equation}}
\renewcommand{\thefigure}{S\arabic{figure}}
\renewcommand{\bibnumfmt}[1]{[S#1]}
\renewcommand{\citenumfont}[1]{S#1}

In the limit where a diatomic molecule can be treated as a point-like dipole, the Rydberg atom-molecule interaction potential is dominated by the charge-dipole interaction $\hat{V}=-\vec{d}\cdot \vec{E}^{\text{ele}}$ where $\vec{d}$ is the dipole moment operator for the molecule and $\vec{E}^{\text{ele}}$ is the electric field of the Rydberg electron at the location of the molecule.  In the case of a $\Lambda$-doublet molecule such as CH, the dipole moment operator is simply a $2\times 2$ matrix which represents the coupling of two energetically nearby states of opposite parity that can be coupled by the presence of an external field, and have been studied extensively in previous work \cite{Rittenhouse2010,Rittenhouse2011,Mayle2012}.  In the case of a rigid-rotor like molecule (such as YbOH), the dipole moment operator is expanded in spherical harmonics in the molecular coordinate.

The Rydberg electron electric field is expanded in the Basis of Rydberg states \cite{GonzalezFerez2015,Kuznetsova2016} with a quantization axis along the along the inter-molecular axis giving
\begin{align}
E^{\text{ele}}_{z,n'l',nl} =& e\sum_{l''=|l-l'|}^{l+l'}\sqrt{(2l+1)(2l'+1)}\begin{pmatrix}
  l' & l'' & l \\
  0 & 0 & 0 
 \end{pmatrix}^2\\
 &\times \left(\int_0^R \frac{r^{l''}}{R^{l''+2}} \psi^*_{n'l'}(r)\psi_{nl}(r)dr+\int_R^{\infty} \frac{R^{l''-1}}{r^{l''+2}} \psi^*_{n'l'}(r)\psi_{nl}(r)dr\right)\nonumber
\end{align}
where $\binom{\cdot\text{~}\cdot\text{~}\cdot}{\cdot\text{~}\cdot\text{~}\cdot}$ is a $3j$-symbol and $\psi_{nl}$ is a radial Rydberg wavefunction for principle quantum number $n$ and angular momentum $l$.  This form of the matrix element is found through a multipole expansion of the electron electric field.  In using this expansion, it is assumed that the molecule is a perfect dipole, with no short-range structure.  The contribution of the short-range electron-molecule interaction can be approximated using a contact interaction with a scattering length set by the electron affinity of the molecule \cite{GonzalezFerez2021}.

As the dipole moment of the molecule increases, an increasing number of Rydberg states are admixed into the electronic state of the system allowing for the Rydberg electron to become more localized.  The larger the dipole moment, the the larger then number of Rydberg states that are mixed, further localizing the electron and deepening the Born-Oppenheimer potentials.  As the dipole moment approaches the Fermi-Teller critical dipole, $d\sim 1.6$ Debye \cite{Turner1977}, the number of Rydberg states needed to converge the potentials sharply increases (as seen in \cite{Mayle2012}).  When the dipole moment exceeds the critical value the electron-dipole binding energy diverges, and a short-range repulsive cutoff is necessary to cut-off this unphysical behavior.  The inclusion of such a cut-off is beyond the scope of this work.  However, we can speculate that the potential depths in such a system would be limited by the depth of the most weekly bound, long-range, electronic state of the dipole-bound molecular anion producing potentials with depth at least in the 10-100 GHz range, ample to produce the large elastic scattering cross section needed for the cooling method described in this work.

We also note that the method in ref.~\cite{Giannakeas2020} may be used to estimate the order of magnitude of the interaction potential depth by modeling the molecule as a perturber that dresses the Rydberg electron and acquires an effective charge. This method may be useful for molecules with supercritical dipole moments.

\end{document}